%% file: main.tex
\documentclass[conference, a4paper]{IEEEtran}

\IEEEoverridecommandlockouts                     \usepackage{booktabs}
\usepackage{graphicx}
\usepackage{multirow}
\usepackage{amsmath,amssymb,latexsym}
\usepackage{rotating}
\usepackage{lettrine}
\usepackage{textcomp}
\usepackage{units}
\usepackage[table]{xcolor}

\usepackage{hyperref}
\usepackage{lipsum}%
\usepackage[pscoord]{eso-pic}
\newcommand{\placetextbox}[3]{%
  \setbox0=\hbox{#3}
  \AddToShipoutPictureFG*{
    \put(\LenToUnit{#1\paperwidth},\LenToUnit{#2\paperheight}){\vtop{{\null}\makebox[0pt][c]{#3}}}%
  }%
}%

\usepackage{tikz}

\usepackage{xcolor}

\title{A Software Platform for Testing Multi-Link Operation in Industrial \mbox{Wi-Fi} Networks
\thanks{This work has been partially funded by SoBigData.it ``SoBigData.it receives funding from European Union – NextGenerationEU – National Recovery and Resilience Plan (Piano Nazionale di Ripresa e Resilienza, PNRR) – Project: “SoBigData.it – Strengthening the Italian RI for Social Mining and Big Data Analytics” – Prot. IR0000013 – Avviso n. 3264 del 28/12/2021.'', and partially funded by the European Commission Horizon Europe SNS JU PREDICT-6G (GA 101095890) Project.}
}

\author{
    \IEEEauthorblockN{
    Matteo Rosani\IEEEauthorrefmark{1}\IEEEauthorrefmark{2}\href{https://orcid.org/0009-0004-4416-4303}{\includegraphics[scale=0.65]{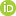}},
    Gianluca Cena\IEEEauthorrefmark{1}\href{https://orcid.org/0000-0003-0084-5321}{\includegraphics[scale=0.65]{orcid_16x16.png}},
    Dave Cavalcanti\IEEEauthorrefmark{3}\href{https://orcid.org/0000-0002-8613-4602}{\includegraphics[scale=0.65]{orcid_16x16.png}},\\ Valerio Frascolla\IEEEauthorrefmark{4}\href{https://orcid.org/0000-0002-4256-2955}{\includegraphics[scale=0.65]{orcid_16x16.png}},
    Guido Marchetto\IEEEauthorrefmark{2}\href{https://orcid.org/0000-0003-3588-9367}{\includegraphics[scale=0.65]{orcid_16x16.png}}, and Stefano Scanzio\IEEEauthorrefmark{1}\href{https://orcid.org/0000-0001-7643-2342}{\includegraphics[scale=0.65]{orcid_16x16.png}}}
    \IEEEauthorblockA{\IEEEauthorrefmark{1}National Research Council of Italy (CNR--IEIIT), Italy. \IEEEauthorrefmark{2}Politecnico di Torino, Italy.}    
    \IEEEauthorblockA{\IEEEauthorrefmark{3}Intel Labs, Intel Corporation, Hillsboro, OR, USA. \IEEEauthorrefmark{4}Intel Labs, Intel Deutschland, Neubiberg, Germany.}
    Email: name.surname@\{cnr.it, intel.com, polito.it\}
    }

\begin{document}
\placetextbox{0.5}{1}{This is the author's version of an article that has been published.}
\placetextbox{0.5}{0.985}{Changes were made to this version by the publisher prior to publication.}
\placetextbox{0.5}{0.97}{The final version of record is available at \href{https://doi.org/10.1109/WFCS60972.2024.10540967}{https://doi.org/10.1109/WFCS60972.2024.10540967}}%
\placetextbox{0.5}{0.05}{Copyright (c) 2024 IEEE. Personal use is permitted.}
\placetextbox{0.5}{0.035}{For any other purposes, permission must be obtained from the IEEE by emailing pubs-permissions@ieee.org.}%

\maketitle
\thispagestyle{empty}
\pagestyle{empty}

\begin{abstract}
Multi-Link Operation (MLO) in \mbox{Wi-Fi 7} is expected to tangibly boost throughput while lowering transmission latency at the same time.
This is very relevant in industrial scenarios and makes MLO suitable, e.g., to support seamless device mobility.
Benefits depend on the ability of multi-link devices to select at run-time the best link, among the available ones, in order to maximize both communication performance and reliability.

In this paper an experimental platform is proposed, with the aim of leveraging commercial hardware and open source software, and easing prototyping and evaluation of MLO techniques.
The platform has been employed to analyze the transmission quality of two pairs of non-overlapping channels, and in particular to assess whether or not adequate diversity is provided, so that those channels can be exploited to improve reliability.
Results point out that correlation between different links is, in most cases, limited, which makes MLO a valuable approach.
\end{abstract}


\section{Introduction}
\label{sec:introduction}

The IEEE 802.11be amendment, best known as \mbox{Wi-Fi} 7, is conceived to provide users with extremely high throughput (EHT), and promises to further reduce the performance gap between wireless LANs and their wired Ethernet-based counterparts.
Not only \mbox{Wi-Fi} \cite{IEEE802.11-20} does not require any fees, but it also retains full compatibility with Ethernet at the data link layer, since both technologies rely on the same addressing space (6B Media Access Control (MAC) addresses, also known as IEEE EUI-48) and have similarly-sized Maximum Transmission Units (MTU).
When network coverage has to be provided over small-to-medium areas (especially indoor)
to support, e.g., mobility of people and devices in enterprises and industrial plants, it often has a clear advantage over cellular networks like 5G.
It must be remarked that next-generation Autonomous Mobile Robots (AMR) will need permanent and high-quality digital data connections, both to coordinate their actions in real-time and to fully integrate in the overall automated production process.
There are great expectations that these advancements can be enabled, at least in part, by \mbox{Wi-Fi} 7 connectivity \cite{frascolla2023}.

Besides the improvements brought to the physical layer and aimed at increasing throughput,
Multi-Link Operation (MLO) is probably the single most prominent feature introduced by \mbox{Wi-Fi} 7.
By and large, it brings the concept of link aggregation to \mbox{Wi-Fi}, by allowing the concurrent use of more than one link.
As of today, simultaneous dual/tri-band operations are available in Access Points (AP), but they are not on end stations (STA).
The main difference between MLO and aggregation techniques exploited on wires, like IEEE 802.3ad, is inherent to the behavior of the underlying communication support.
In the case of Ethernet, as long as cables are not defective, the probability that a frame is corrupted while traveling on a link is mostly negligible.
This is not true at all in \mbox{Wi-Fi} (and more in general for any wireless transmission technologies).
In fact, in typical operating conditions, the likelihood that a transmission attempt fails (that is, that the frame is not delivered to destination successfully) can not be ignored.
Even worse, channel conditions may vary noticeably and abruptly, as a consequence of, e.g., 
device mobility, obstacles and reflections, interference from nearby wireless devices, and electromagnetic noise produced by power equipment like industrial machinery\cite{Merwaday2023}.
For this reason acknowledged transmissions are customarily adopted in \mbox{Wi-Fi} for unicast traffic, which enable Automatic Repeat reQuest (ARQ) mechanisms.
While retransmissions ensure that a packet is eventually delivered most of the times, they may sensibly increase both latency and jitters, which negatively impact determinism.
For this reason, wireless networks are seldom employed in real-time control applications.
MLO is expected to be a game-changer in this scenario.
By allowing the link used for frame transmission to be chosen at run-time (among a small selection of frequencies, usually chosen in different bands to improve diversity),
a better and more stable behavior can be achieved, also leading to shorter latency.

Basically, a Multi-Link Device (MLD) as defined in \mbox{Wi-Fi} 7 consists of a number of distinct affiliated STAs (usually two or three) termed \mbox{L-MACs}, each one tuned on a distinct channel
and associated to an \mbox{L-MAC} of the AP.
Operations of \mbox{L-MACs} in an MLD STA are coordinated by a single entity called the \mbox{U-MAC}.
It is worth pointing out that, unlike channel bonding, contention and access to the involved wireless channels are performed by \mbox{L-MACs} in a totally independent way.
Concerning the \mbox{U-MAC}, its roles include managing the transmission buffer in a unified manner and deciding, for every one of the enqueued frames, which \mbox{L-MAC} (and hence, channel) has to be used.
A fair amount of work is (still) needed to analyze how \mbox{U-MAC} operation impacts on the overall communication quality (considering, e.g., throughput, latency, jitters, losses, and power consumption)
and to define suitable algorithms able to optimize performance indicators depending on the specific application context.
Moreover, there is the feeling that \mbox{Wi-Fi 8} \cite{giordano2023wifi} (under development as part of the IEEE 802.11bn Task Group, whose activities have just started),
which is meant to provide ultra high reliability (UHR) and is hence of particular interest for industry, will also ground on MLO by defining suitable extensions, e.g., distributed MLO.

In this paper a low-cost experimental MLD architecture is described, called Virtual Multi-Link Device (VMLD) and based on commercially available PC hardware and open source operating system (OS).
It can help performing research activities by providing an experimental platform that exposes MLO capabilities in user space, where it is easier to apply modifications than in kernel space.
Similar platforms exist, like the one described in \cite{9417329}. 
The authors of this paper designed a high-rate monitoring platform capable of extracting crucial information from the device driver and transfer it to user space via polling, events, or an hybrid approach.
Another possibility to test MLD capabilities in a dense deployment is to use an emulator like in \cite{CAPDEHOURAT20181}.
In both cases, some modifications are needed to make a fully working solution.

The paper is structured as follows:
in Section~\ref{sec:VMLD} the VMLD architecture is introduced in general terms,
while details about its practical implementation are given in Section~\ref{sec:implementation}
In Section~\ref{sec:results} an experimental campaign is described where two VMLDs have been exploited to study the quality of links operated in parallel, and results about losses and latency are provided.
Finally, some conclusions are drawn in Section~\ref{sec:conclusion}.

\section{Virtual MLD}
\label{sec:VMLD}
The proposed VMLD platform is mainly designed to enable rapid prototyping and testing of MLO techniques in software,
especially on the STA side. 
Besides, it also permits
to acquire logs about the overall MLD behavior, which include information (outcomes, timings, etc.) on every transmission performed on air by its \mbox{L-MACs}.
Such logs constitute a sort of big data on which post-analysis algorithms and machine learning techniques can be then applied \cite{10295470,s22134925,https://doi.org/10.1002/itl2.326,9786784}. 
To this aim, a number of specific functions and mechanisms have been purposely defined.
Datasets collected from experimentation permit complex scheduling techniques conceived for the \mbox{U-MAC} to be analyzed in a simplified software environment, before being integrated into real devices. 
Sensible examples are, e.g., procedures for managing device roaming \cite{10144124} and for best-channel selection \cite{9152055}.

A possible final target for the proposed architecture foresees a software implementation of MLO in user space on a UNIX/Linux platform. 
The availability of such a platform enables easier and faster development, experimental verification, and comparison of MLO techniques, in order to determine the best proposals to be then implemented at the driver level or directly in the network adapter firmware. 
For example, it is worth remembering that floating point arithmetic is typically unavailable at the kernel level, which makes impractical implementing there some classes of algorithms.
Working at the user level unavoidably increases processing latency and jitters due to the necessary interactions with the OS (system calls, context switches, ...).

The VMLD platform can be exploited to reliably assess the performance and correctness of MLO algorithms 
(e.g., mechanisms for dynamically dispatching packets to \mbox{L-MACs}) in user space,
but OS latency, especially those concerning context-switching operations, must be carefully evaluated and taken into account, as in \cite{6489584}. 
Techniques for improving determinism of UNIX/Linux systems, such as using and configuring real-time kernel extensions like PREEMPT\_RT, are described in \cite{8477080,9756165,5637989}. 
Performing an accurate timing analysis of the different sources of internal interference for the entire VMLD platform 
and their contributions to the overall latency
is beyond the scope of this paper, which focuses instead on describing this architecture from a functional point of view.
Some sample datasets acquired with it will be analyzed to assess overall communication performance, mainly considering aspects related to channel access.

\begin{figure}[t]
    \begin{center}
    \includegraphics[width=1.0\columnwidth]{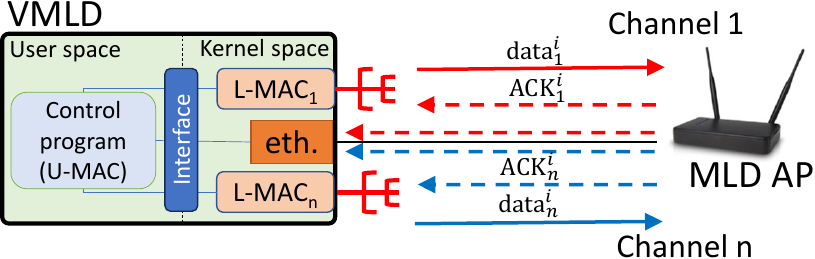}
    \end{center}
    \vspace{-0.2cm}
    \caption{General schema of the VMLD architecture.}
    \vspace{-0.3cm}
    \label{fig:arch1}
\end{figure}

The schematic diagram of the proposed architecture is sketched in Fig.~\ref{fig:arch1}, 
where a commercial PC takes the role of a VMLD device.
A modular architecture was adopted, where some components can be disabled or enabled to selectively activate specific features.
From the software point of view, the number of \mbox{L-MACs} is configurable (concerning the hardware, our equipment currently supports two PCI Express network boards). 
A \textit{control program} running in user space manages the transmission of frames and reacts to direct/indirect frame reception events.
Generally speaking, it implements the part of the VMLD (including the \mbox{U-MAC}) we are mostly interested in.
In the specific case of this paper, the control program was exploited for acquiring statistics about transmissions on air,
so as to provide some figures about the real performance one may expect from the platform.

To this purpose, information (e.g., outcome and delivery time) must be acquired for every packet $i$ sent on the \mbox{L-MAC} tuned on any specific channel $ch$ (we denote data$^i_{ch}$).
As shown in Fig.~\ref{fig:arch1}, this can be practically obtained in two distinct ways: 
either by intercepting the reception of the acknowledgment (ACK) frame on the related wireless adapter, 
or via a wired Ethernet connection that forwards the packet from the AP back to the originating PC (device ``eth.'' in the figure).
In the latter case, which serves for detecting correctly delivered packets for which the ACK frame went lost,
the destination MAC address must be set equal to the address of the local Ethernet board.
In both cases the transmission/reception timestamps are taken using the same time base,
which means that a clock synchronization mechanism is not needed on the VMLD platform
to enable precise 
latency measurements \cite{9524299,6489696,9169833}.

\begin{figure*}[t]
    \begin{center}
    \includegraphics[width=2.0\columnwidth]{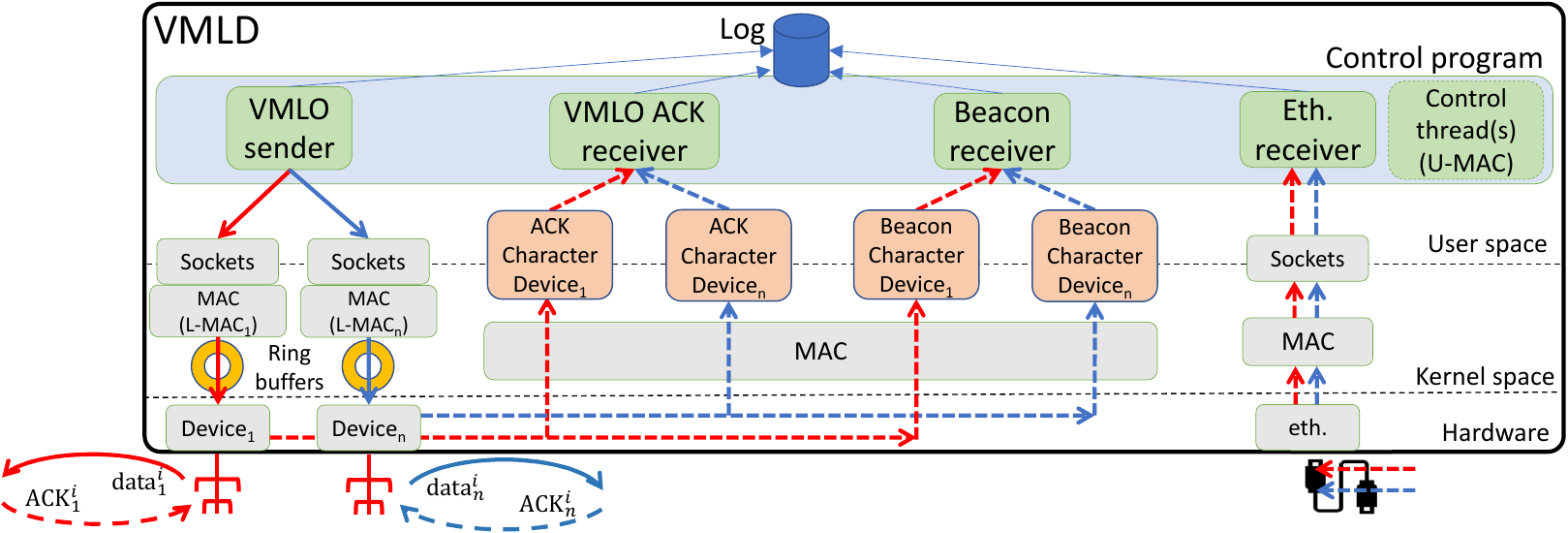}
    \end{center}
    \vspace{-0.2cm}
    \caption{Detailed scheme of the VMLD architecture 
    (including the sender path as well as the ACK, beacon, and Ethernet receiver paths).}
    \vspace{-0.3cm}
    \label{fig:arch2}
\end{figure*}

\section{Implementation}
\label{sec:implementation}
The detailed diagram in Fig.~\ref{fig:arch2} depicts all the main building blocks of the proposed VMLD architecture. 

\subsection{Transmission path}
The left side of the figure describes the sending path, managed by the \textit{VMLO sender} thread. 
In the current implementation, this thread uses the \texttt{sendto()} function made available by POSIX sockets to enqueue packets in the $n$ available wireless interfaces (termed Device$_1$,..., Device$_n$), each of which corresponds to a specific \mbox{L-MAC}. 
To avoid saturating CPU cores, only one \textit{VMLO sender} thread was created. 
Since we are using only two threads (one for the sender and one for the receiver), there are not any scalability issues concerning the number of possible \mbox{L-MACs}.
Clearly, other design choices are also possible, depending on the specific MLO features to be emulated.
For instance, more than one thread could be used for implementing the \mbox{\mbox{U-MAC}}, and functions other than \texttt{sendto()} could be exploited for interacting with \mbox{L-MACs}.
It should be noted that some drivers, like the Atheros \texttt{ath} ones we used in this work, are highly customizable, e.g., they permit disabling retransmission and backoff.
This made experimentation with the VMLD easier. 

Concerning the experiments described below, the \mbox{U-MAC} (which is completely implemented in user space) feeds every single packet on all the \mbox{\mbox{L-MACs}}, hence modeling MLO usage to achieve highly-reliable redundant communication.
In the context of the experiments described in this paper, the order with which enqueuing takes place on the different interfaces is selected randomly, to avoid biases in computing statistics. 
Another opportunity given by the VMLD platform could be, e.g., to test traffic steering policies implemented in user space, where the \mbox{L-MAC} is selected on a per-packet basis depending on the current quality of channels.
Doing so requires the transmission ring buffer to be emulated in software as part of the \mbox{U-MAC}.
More complex, but likely feasible, are procedures that transmit every packet on all the \mbox{L-MACs}, but which avoid enqueuing a packet on an \mbox{L-MAC} when an ACK frame has already been received for it on at least another \mbox{L-MAC}.
Doing so permits to save bandwidth while ensuring high dependability.
Probably, the most interesting target for the VMLD platform are custom algorithms that decide at run-time how many and which \mbox{L-MACs} to use for transmitting every single packet, to guarantee a certain quality of service.

\subsection{Acknowledgment path}
Notifications about the outcomes of transmissions (success/failure) are managed by the \textit{VMLO ACK receiver} thread. 
It relies on the \texttt{poll} system call to check the content of $n$ character devices (one per \mbox{L-MAC}),
which are used to transfer \textit{ACK} events and other relevant statistical information 
from the kernel to the user space. 
The same mechanism is also used to convey \textit{ACKtimeout} events, which are generated when all the possible retries of a given packet have been carried out without receiving any ACK frame (our \mbox{Wi-Fi} boards do not provide any interrupts on timeouts of intermediate retries).
Every character device is fed by the device driver with information pertaining to the most recent frame transmission performed by the related \mbox{L-MAC}
(e.g., the ACK timestamp and the number of retries actually performed by the network board for any given packet). 
To this purpose, the device driver had to be modified to write this information in the character device every time an ACK or ACKtimeout event is detected on one of the $n$ wireless network interfaces. 
Identifying the correct position where the code for managing these events must be inserted in the specific driver is one of the most complex aspects that must be analyzed for implementing this function in the VMLD.

In the context of SoftMAC drivers \cite{10.1145/3349623.3355477}, like the \texttt{ath} driver series used in this work,  where a large portion of the MAC is implemented in software, detection of these events is done in the \texttt{ieee80211\_tx\_status()} function\footnote{The \texttt{ieee80211\_tx\_status()} function is located in the \texttt{net\slash mac80211\slash status.c} file. 
To trigger the execution of this function for every received frame, the \texttt{MAC80211\_DEBUGFS} option must be activated during kernel compilation, and the attribute \texttt{\slash sys\slash kernel\slash debug\slash ieee80211\slash phy0\slash force\_tx\_status} of \textit{debugfs} must be set to \texttt{1}. 
The term \texttt{phy0} is a reference to the \mbox{Wi-Fi} adapter.}. 
We decided to bring the modifications required by the VMLD architecture in this particular position of the code because it fits in the part of the driver that is independent of both the specific wireless adapter and the specific version of the IEEE 802.11 standards it relies on.
Together with one of the two possible events (either ACK or ACKtimeout) the driver transfers in the relevant character device part of the \texttt{ieee80211\_tx\_info} data structure, which contains the transmission status.
Among the included information there are, for instance, the number of transmission attempts for every transmission series, the corresponding transmission rate, and the received signal strength indicator (RSSI) of the ACK frame. 
The same data structure also includes device-dependent parts, which currently are not copied in the character device because they are too specific (that is, they are primarily related to the device and its \mbox{Wi-Fi} version).
The case of FullMAC drivers, where the entire MAC layer is implemented in the hardware (or firmware) of wireless adapters, is much more complex. 
In this case, very few information is made available about transmitted frames. 
For this reason, the analysis of FullMAC drivers is left as future work.

Regarding the \textit{VMLO ACK receiver} thread in user space, upon reception of information about 
ACK/ACKtimeout events the \texttt{wait\_event\_interruptible()} macro is used to raise a \texttt{POLLIN} event and notify the presence of data in the character device. 
This event unblocks the \texttt{poll()} system call, allowing the user space thread to read in and suitably deal with the relevant data associated with the frame (in the current implementation they are simply logged for post-analysis). 
Another option to implement synchronization on the acknowledgment path is to generate a number of threads equal to the number of character devices, each one using the \texttt{read()} system call to wait for the insertion of data into the character device. 
Since the two above solutions are not expected to show substantial differences in terms of the added jitters (contributions related to the context-switch between kernel and user space are predominant in this case), 
we decided to rely on the one based on the \texttt{poll()} system call, which avoids generating too many threads, thus saturating the available cores.

\subsection{Additional functions}
The \textit{beacon receiver} thread is also based on character devices and its structure closely resembles the \textit{VMLO ACK receiver} thread.
It allows statistics related to beacon frame reception to be transferred to the program in user space. 
In this way, the information contained in beacons, or derived from their reception, like the related RSSI value and reception time, is made available to the whole VMLD architecture (including the \mbox{U-MAC}) and can be exploited to drive decisions, for example, to facilitate the roaming from one AP to another.

In addition, the VMLD architecture also provides the ability to detect notifications when a frame is received on its  Ethernet interface that was sent by one of its wireless interfaces (kind of a loopback that also includes the \mbox{Wi-Fi} link). 
This is trivially achieved by using the typical \texttt{recv()} function (system call) of sockets (see the \textit{eth. receiver} thread on the right side of Fig.~\ref{fig:arch2}). 
This peculiar path permits to acquire reliable statistics about transmission outcomes, which are not affected by the loss of ACK frames (the probability that the frame is lost in the return path on wires, once it has been received by the AP, can be quietly neglected, while the probability that the ACK frame sent on air by the AP is corrupted cannot).

These last two types of threads have yet to be implemented in a stable form, because this paper primarily focuses on those essential features that are strictly needed for implementing MLO in a VMLD (notably, these are the most critical functions to implement and test). 
Finally, \textit{control} threads manage the execution of all the other previously described threads.

\section{Results}
\label{sec:results}
This section describes the experimentation we carried out to test some basic functions of the proposed VMLD architecture, 
and also discusses the configuration of several specific parameters of the device driver for tuning its behavior so as to meet a number of requirements.
To show the possibilities offered by the platform, we used it to perform a thorough analysis of the quality of a number of channels at the same time, by acquiring suitable information about transmissions performed in a real environment.
As said before, this corresponds to using MLO to support seamless redundancy in \mbox{Wi-Fi} (the \mbox{U-MAC} simply replicates every frame on all \mbox{L-MACs}).
The following subsection provides details on the dataset acquisition using VMLD,
while Subsections~\ref{sub:FDR} and~\ref{sub:latencies} analyze the link from the point of view of the frame delivery ratio (FDR) and transmission latency, respectively.

\subsection{Dataset acquisition}
\label{sub:database}
For dataset acquisition, the \textit{VMLO sender} thread was programmed to cyclically generate packets, whose size is $\unit[50]{B}$, every $\unit[0.5]{s}$.
In the experiments below, this is done to ensure that adjacent transmission attempts on the same channel are independent.
Every \mbox{L-MAC} encapsulates the packet in a \mbox{Wi-Fi} frame and sends it to the AP to which it is associated.
Since in the context of this paper we employ the proposed VMLD architecture to analyze the quality of (contextual) wireless links, automatic frame retransmission was disabled (i.e., every frame is sent only once), 
the backoff procedure was disabled (i.e., the value used for backoff is always $0$), 
and the bit rate was set to a fixed value ($\unit[54]{Mbps}$). 
In this way, link quality was determined at regular intervals, evenly spaced in time,
without all the uncertainties typically introduced by the MAC mechanisms of \mbox{Wi-Fi},
such as the variable number of transmission attempts, the random duration of backoff, 
and the dynamic selection of the transmission speed, as specified by the modulation and coding scheme (MCS).
Another reason to configure parameters this way, which is not directly related to the MLO, is to show the practical possibilities offered by the architecture. 
For example, it permits frame retransmission and packet queues to be handled directly in user space (with some limitations), e.g., within the \textit{control} thread.

To check the impact on performance of the computing platform and the wireless spectrum, two VMLDs were configured on two distinct Linux PCs, 
with different CPUs and motherboards but using the same kernel (version 6.2.0-36). 
Each PC is equipped with two TP-Link TL-WDN4800 \mbox{Wi-Fi} network adapters that virtually implement two \mbox{L-MACs}. 
Both are managed by the \texttt{ath9k} device driver version \texttt{linux-source-6.2.0}. 
By doing so, we managed to perform the experimental analysis on four distinct non-overlapping channels in the $\unit[2.4]{GHz}$ band at the same time.
To prevent mutual interference, we selected channels $\langle 1,5 \rangle$ on the first VMLD and $\langle 9,13 \rangle$ on the second  (in the following termed ch1, ch5, ch9, and ch13).
Both PCs were located in our laboratory, spaced by a few meters so that they can not interfere with each other due to bleeding phenomena \cite{8060997}.
It is important to point out that the presence of many nearby interfering nodes operating in the same band makes the environment very challenging (and of practical interest when dense indoor environments are considered, like industrial plants).
To provide a glimpse on this point, during the acquisition of the dataset there were 
$9$ APs configured on ch1, one on ch3, one on ch5, $5$ on ch6, $3$ on ch9, $6$ on ch11, and one on ch13. 
Although the dataset was acquired in a laboratory, previous experiments have shown that, concerning interference, it is actually very similar to a real industrial environment \cite{ExperimentalEvaluationSeamless2017}. 

The sender node logs the outcome $x_i$ of every packet $i$ in the dataset.
In particular, the ACK event denotes a correctly received packet ($x_i=1$), while the ACKtimeout event is interpreted as a non-received packet ($x_i=0$). 
Since ACK frames may sometimes be lost, this leads to slightly overestimated losses.
However, doing so provides the correct probability for a frame to be retransmitted.
The transmission latency $d_i=\frac{t_\mathrm{S}-t_\mathrm{R}}{f_{\mathrm{TSC}}}$ of packet $i$, also stored in the dataset, was obtained by computing the difference between the timestamp $t_\mathrm{S}$ acquired in user space by the \textit{VMLO sender} thread just before invoking the function \texttt{sendto()} and the timestamp $t_\mathrm{R}$ acquired in kernel space within the function \texttt{ieee80211\_tx\_status()}.
Doing so is slightly optimistic, because the (small) notification time from kernel to user space is not included. 
Timestamps were obtained by reading the timestamp counter (TSC) register of the CPU, which in our configuration is increased by one every base clock cycle (with frequency $f_{\mathrm{TSC}}$). 
The TSC provides a very precise and stable time base and suffers from very limited jitters because it does not involve any context switches.
Frequency scaling was disabled to obtain accurate time measurements. 
This method permits taking timestamps without the need to perform any context-switch operations or call system calls, which noticeably lowers jitters and provides precise results.
The acquired dataset includes $N=1,261,725$ samples, which corresponds to more than one week of continuous VMLD operations and enables performance to be reliably assessed.

\subsection{Frame delivery ratio}
\label{sub:FDR}
\begin{figure}[t]
    \begin{center}
    \includegraphics[width=1\columnwidth]{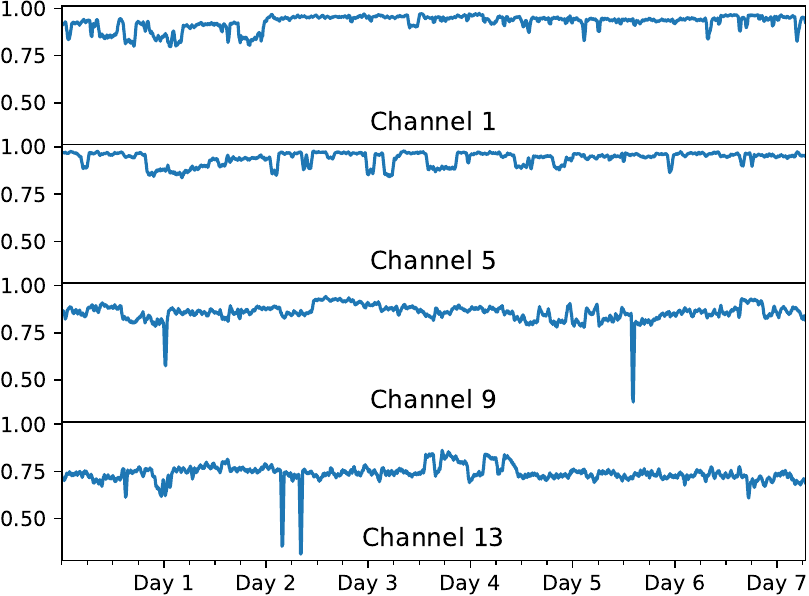}
    \end{center}
    \vspace{-0.2cm}
    \caption{FDR computed on a moving window of width 30 minutes for the four analyzed channels (every tick in the x-axis is 6 hours).}
    \vspace{-0.3cm}
	\label{fig:fdr}
\end{figure}

The acquired dataset was firstly analyzed in terms of the FDR, which is defined as the fraction of frames correctly delivered to the recipient over the frames sent in a given time interval.
The FDR, which only depends on the conditions of the wireless channel (no transmission attempts could be dropped because of the VMLD internal architecture) permits to assess communication reliability.
As said before, the number of frames that arrive to the destination is inferred indirectly by counting the number of ACK frames correctly received.

In time diagrams of Fig.~\ref{fig:fdr}, a simple moving average was employed to evaluate the FDR on the four analyzed channels on 3,600 adjacent samples, corresponding to $30$ minutes.
Reducing the width of the window makes plots more irregular, since transmission on a wireless channel is basically a random binary process.
As can be seen, ch1 and ch5 are characterized by more frequent variations of the FDR, which lie in the range from $79\%$ to $98\%$ percent.
Conversely, ch9 and ch13 were seemingly more stable, although some peaks are present where the FDR drops down to about $30\%$.

A first consideration is that, channel conditions are very dynamic, and every channel differs from the others. 
Consequently, MLO where the \mbox{U-MAC} relies on best channel selection 
(or the simultaneous transmission of the same frame on multiple channels) could be really helpful for industrial communications, even when, as in this experimental campaign, all \mbox{L-MACs} are operated in the same $\unit[2.4]{GHz}$ band.
MLO approaches that exploits multiple bands ($\unit[2.4]{GHz}$, $\unit[5.0]{GHz}$, and $\unit[6.0]{GHz}$)
are intuitively characterized by a much better diversity, and the key for facing the erraticness of the wireless spectrum.
The VMLD architecture could be very valuable, e.g., to find the most suitable algorithms for link selection in MLO,
especially in the prototyping phase.

\begin{table}[t]
    \caption{Correlation between TX outcomes ($x_i$) on different channels.}
    \label{tab:corr_outcomes}
    \centering
    \normalsize
    \begin{tabular}{c|cccc}
    & ch1 & ch5 & ch9 & ch13 \\
\hline
ch1 &  \cellcolor{gray!40}1.000 & \cellcolor{gray!20}0.021 & 0.007 & 0.000 \\
ch5 & \cellcolor{gray!20}0.021 & \cellcolor{gray!40}1.000 & 0.004 & -0.004 \\
ch9 & 0.007 & 0.004 & \cellcolor{gray!40}1.000 & \cellcolor{gray!20}-0.113 \\
ch13 & 0.000 & -0.004 & \cellcolor{gray!20}-0.113 & \cellcolor{gray!40}1.000 \\
    \end{tabular}
    \vspace{-0.3cm}
\end{table}

Table~\ref{tab:corr_outcomes} reports the  Pearson correlation coefficient between the outcomes of different channels. 
In particular, the correlation between two channels a and b is computed as:
\begin{equation}
    \rho_{\mathrm{a},\mathrm{b}} = \frac{\sum_{i=1}^{N} (x^{\mathrm{a}}_i-\mu_{\mathrm{a}})(x^{\mathrm{b}}_i-\mu_{\mathrm{b}})}{\sqrt{\sum_{i=1}^{N} (x^{\mathrm{a}}_i-\mu_{\mathrm{a}})^2}{\sqrt{\sum_{i=1}^{N} (x^{\mathrm{b}}_i-\mu_{\mathrm{b}})^2}}}
\label{eq:correlation}
\end{equation}
where $x_i^{\mathrm{a}}$ and $x_i^{\mathrm{b}}$ represent the outcomes of the transmission of packet $i$ on channels a and b, while $\mu_{\mathrm{a}}$ and $\mu_{\mathrm{b}}$ represent the mean values of outcomes (i.e., the FDR 
of the two channels evaluated over the entire dataset).

As can be seen, there is a practically negligible correlation between ch1 and ch5 and a significant negative correlation between ch9 and ch13. 
Relevant cases in the table have been highlighted with a light grey color. 
Negative correlation in the second case could be due to the adjacent channel interference (ACI) phenomenon \cite{8060997, 6213952}.
In fact, the two adapters tuned on ch9 and ch13 are located in the same PC and fastened to adjacent PCI express slots, which generates interference (signal bleeding) even between non-overlapping channels. 
ACI typically affects negatively the clear channel assessment (CCA) mechanism, and rarely they can corrupt outgoing frames. 
In the first case, the sending of a frame on one channel may generate collisions on the other one, generating a negative correlation as happens in this context.

\subsection{Transmission latency}
\label{sub:latencies}

\begin{figure}[b]
    \begin{center}
    \vspace{-0.3cm}
    \includegraphics[width=1\columnwidth]{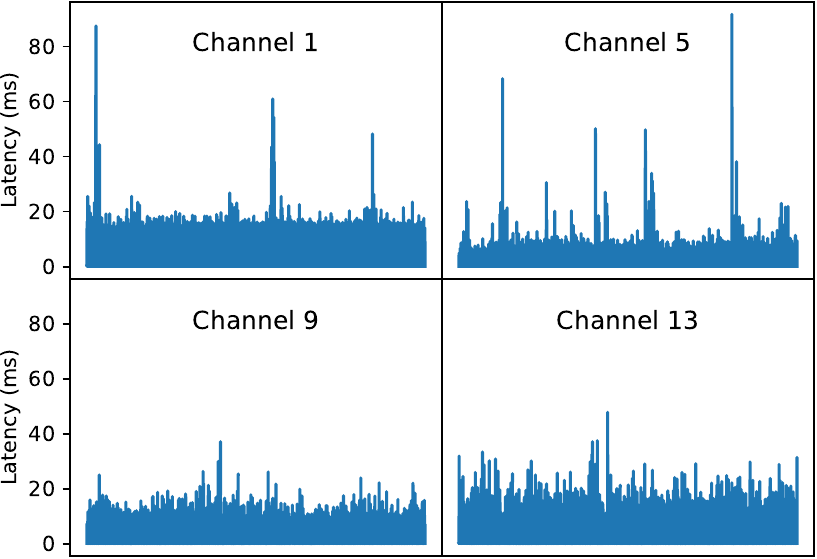}
    \end{center}
    \vspace{-0.2cm}
    \caption{Transmission latency vs. time for the four analyzed channels.}
	\label{fig:latency}
\end{figure}

The four channels exploited by the two VMLDs were also analyzed from the point of view of latency. 
Plots in Fig.~\ref{fig:latency} show timing diagrams of the latency $d_i$ of all the transmitted packets. 
As for transmission outcomes, this quantity is not stationary, and every channel exhibits a different behavior.

Latency provides a direct indication of the network load generated by the nearby nodes on the frequency used by the link (including partial overlaps), since in \mbox{Wi-Fi} a node is only allowed to start transmission when it senses the channel idle.
Since we are interested in the behavior of the different links of an MLD, the same correlation analysis performed for the outcomes was repeated for latency using \eqref{eq:correlation}, where outcomes $x_i$ were replaced by latency $d_i$.

Results, reported in Table~\ref{tab:corr_latencies}, are quite interesting.
Unlike frame losses, overall latency may include contributions due to both internal delays introduced by the processing platform and waiting times related to channel access.
As expected, interference generated by nodes tuned on channels other than $1$, $5$, $9$, and $13$ may create tangible dependencies between the latencies experienced on the two \mbox{L-MACs} of our VMLDs, as witnessed by the high correlation between ch9 and ch13 ($0.386$), which was likely caused (in part) to nodes tuned on channel $11$. 
A non-negligible correlation of $0.042$ was also found between ch5 and ch9, which can not be due to internal delays because the two network interfaces are installed on different PCs. 
In this case, the only explanation is the activity of the APs (five, for our experimental environment) configured on channel $6$.
Since that channel overlaps with both ch5 and ch9, transmissions on it interfere with both VMLDs at the same time, delaying channel access.
However, in this case only one \mbox{L-MAC} is affected per MLD, which does not impair diversity for it.

\begin{table}[t]
    \caption{Correlation between TX latency ($d_i$) on different channels.}
    \label{tab:corr_latencies}
    \centering
    \normalsize
    \begin{tabular}{c|cccc}
    & ch1 & ch5 & ch9 & ch13 \\
\hline
ch1 & \cellcolor{gray!40}1.000 & -0.005 & 0.000 & 0.002 \\
ch5 & -0.005 & \cellcolor{gray!40}1.000 & \cellcolor{gray!20}0.042 & -0.002 \\
ch9 & -0.000 & \cellcolor{gray!20}0.042 & \cellcolor{gray!40}1.000 & \cellcolor{gray!20}0.386 \\
ch13 & -0.002 & -0.002 & \cellcolor{gray!20}0.386 & \cellcolor{gray!40}1.000 \\
    \end{tabular}
    \vspace{-0.3cm}
\end{table}

In the following, we focus on ch1 and ch5, and especially on ch1 because it seemingly does not suffer from 
above phenomena sensibly.
\begin{table*}[t] 
    \caption{Statistics about latency evaluated on all packets, acked packets only, and not acked packets only ($\mu s$).}
    \label{tab:stats_latency}
    \centering
    \resizebox{\textwidth}{!}{\input{tabs/stats_horizontal}}
    \vspace{-0.2cm}
\end{table*}
Table~\ref{tab:stats_latency} reports the main statistical indices related to the latency of ch1 and ch5 evaluated over three different sets of packets: \textit{all} (i.e., set $\{x_1, x_2,...,x_N \}$), \textit{acked} (i.e., set $\{x_i \mid 1 \leq i \leq N, x_i=1 \}$), and \textit{not acked} (i.e., set $\{x_i \mid 1 \leq i \leq N, x_i=0 \}$).
Please note that, for frames for which an ACK was not received, latency refers to the ACKtimeout expiry.

As a first consideration, not acked frames are characterized by a larger standard deviation than the acked ones. 
This is because there may be several conditions under which the transmitter node may not have received the relevant ACK frame correctly: 
1) the transmitted data frame successfully arrived at the receiver node and so the ACK frame was returned, but it arrived corrupted at the receiver (the time when failure is detected may vary depending on interference);
2) the transmitted data frame arrived corrupted at the receiver node and so the ACK frame was not sent, 
causing a timeout in the sender.
In all these cases the network board generates an interrupt after a time interval whose duration depends on the specific condition among those listed above,  which increases the measured standard deviation.

Some statistical indices, like the minimum and maximum, need a large number of samples to be reliably estimated. 
Since the statistics of not acked frames were calculated using only $7.3\%$ and $6.0\%$ of the whole dataset for ch1 and ch5, respectively, they are noticeably less reliable than statistics of acked frames.
The table also shows percentiles. 
In particular, the 99.9-percentile (i.e., $P_{99.9}$) shows that for 99.9\% of the frames sent on ch1 the ACK notification arrives to the sender node within $\unit[13.5]{ms}$, while not acked frames take longer.
The $P_{99.9}$ of ch5 is better, and its value is about $\unit[6.7]{ms}$ for all the acked frames, while for not acked frames it is $\unit[7.2]{ms}$.

For the \mbox{Wi-Fi} networks involved by the VMLD platform we used for experimentation, 
the ACK frame (which lasts $T_{\mathrm{ACK}}=\unit[44]{\mu s}$) is transmitted after a Short Interframe Space (SIFS)  whose duration is $T_{\mathrm{SIFS}}=\unit[16]{\mu s}$.
Consequently, we are certain that the data frame arrived to the AP at least $T_{\mathrm{SIFS}}+T_{\mathrm{ACK}}=\unit[60]{\mu s}$ before the times reported in the table, 
which are obtained by timestamping the interrupts related to ACK and ACKtimeout events.

\begin{figure}[tb]
    \begin{center}
    \includegraphics[width=1\columnwidth]{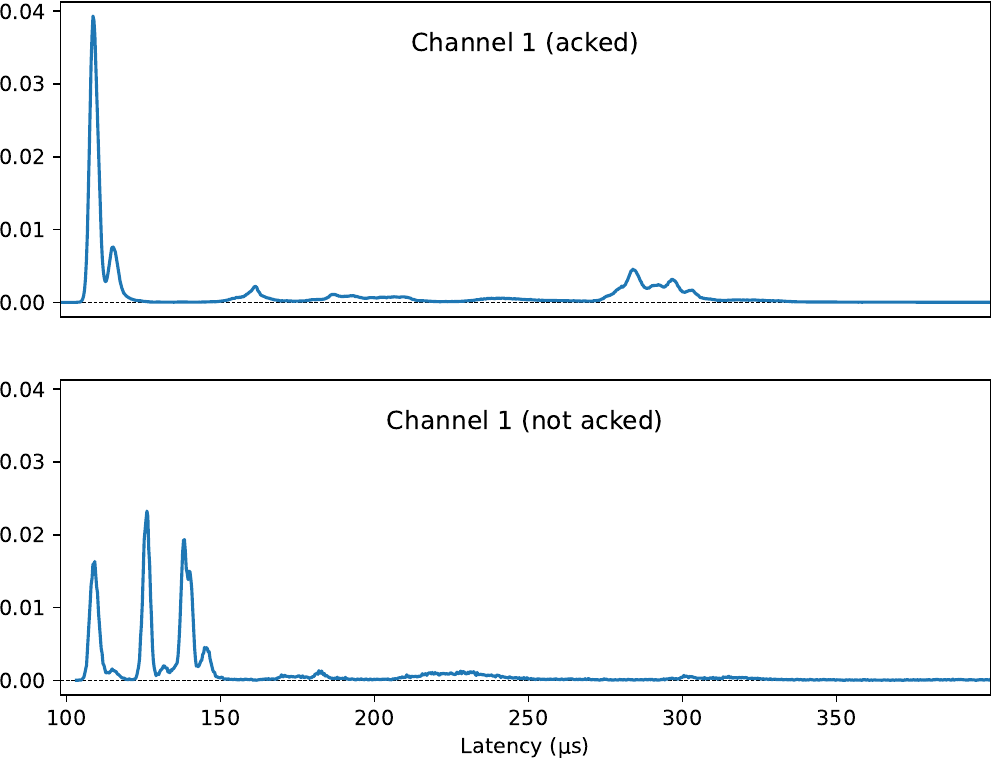}
    \end{center}
    \vspace{-0.3cm}
    \caption{PDF of the latency for ch1 (acked and not acked packets).}
    \vspace{-0.2cm}
    \label{fig:pdf_latency}
\end{figure}

To give additional insights on latency,
the probability density function (PDF) of both acked and not acked packets was calculated, 
and reported in the two plots of Fig.~\ref{fig:pdf_latency}.
Regarding the PDF of acked frames, the first peak corresponds to frames that were sent immediately since they found the channel idle. 
Asynchronous interference due to the channel found busy causes random delays, whose distribution is approximately flat depending on the size of interfering frames.
The other peaks, instead, could be related to the specific situation where the node has a frame ready to be sent, the channel is sensed as busy, and it must wait. 
If the sender, which waits for Distributed Inter Frame Space (DIFS) since its contention window was set to $0$,
competes with a pending beacon, which is sent by the AP after a PCF Interframe Space (PIFS),
the latter has precedence and delays the former for a time that is mostly fixed for every AP.
This explains the presence of peaks in the PDF.
Regarding the PDF of not acked frames, the three peaks are due to the different conditions (listed above) for which an ACK frame may not be received by the sender.
An additional effect that may impact on latency is ACI. 

\begin{figure}[t]
    \begin{center}
    \includegraphics[width=\columnwidth]{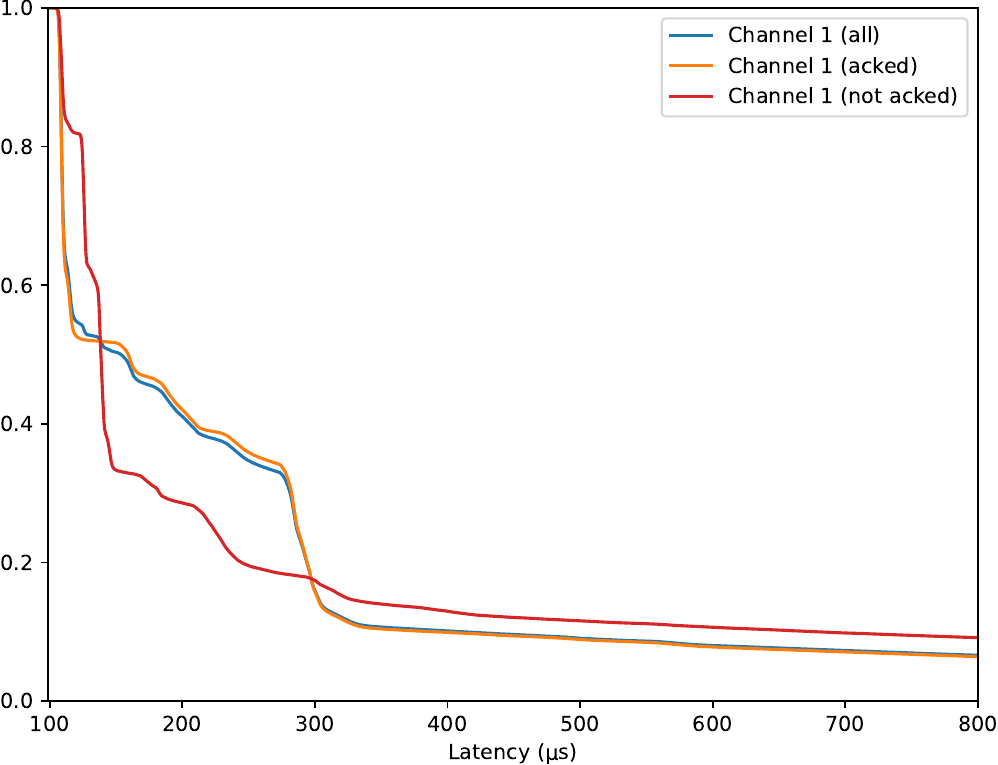}
    \end{center}
    \vspace{-0.3cm}
    \caption{CCDF of the latency for ch1 (acked and not acked packets).}
    \label{fig:ccdf_latency}
    \vspace{-0.2cm}
\end{figure}

Latency of acked, not acked, and all frames is also compared through their complementary cumulative distribution functions (CCDF) reported in Fig.~\ref{fig:ccdf_latency}.
Since the majority of the transmitted frames ($92.7\%$ of the dataset, corresponding to $1,169,619$ samples) is properly acknowledged, the curves related to the all and acked subsets mostly overlap. 
CCDFs show that latency exceeds $\unit[\sim 350]{\mu s}$ for less than about $10$\% of the frames,
and that the distribution of not acked frames is rather different than the one of acked frames.

\section{Conclusion}
\label{sec:conclusion}
By enabling STAs to simultaneously use multiple wireless links, MLO has the potential to bring tangible improvements to \mbox{Wi-Fi} performance and dependability, 
enabling its adoption for time-sensitive applications in industrial scenarios.
To this purpose, the ability of MLD equipment to select at run-time the most suitable link on which packets have to be sent is essential.
This is even more true when seamless redundancy is exploited to provide dependable communication services in spite of the intrinsic erraticness of the wireless spectrum.
Proper MLO implementation requires a new class of mechanisms that permit the MAC level of \mbox{Wi-Fi}
to exploit diversity.

The  Virtual MLD architecture we describe in this paper has been conceived to support research activities concerning these mechanisms, by easing their prototype implementation and experimental assessment.
Basically, it consists of a low-cost platform that relies on widely available legacy wireless equipment (typically complying to \mbox{Wi-Fi} $4$ to $6$) and emulates MLO in software.
This is not particularly limiting, as MLO is primarily meant to improve the basic Enhanced Distributed Channel Access (EDCA).
The ability to implement part of the operations in user space eases code development and debugging, which means that the platform is not expected to become immediately obsolete when commercial boards complying to \mbox{Wi-Fi} 7 with open source drivers will be available.

In this paper, the VMLD platform has been employed to implement a simple redundancy scheme where all packets are sent at the same time on two wireless interfaces.
Then, an experimental campaign has been carried out, which lasted one entire week, where suitable metrics (transmission outcomes and latency) were acquired to analyze the quality of two links tuned on non-overlapping channels and operated contextually.
Results highlight that the two links are not totally independent.
Instead, cross-interference was observed on the wireless medium, due to the activity of nearby nodes tuned on frequencies that overlap with both our links, and bleeding phenomena due to the proximity of the radios of \mbox{L-MACs} in the VMLD.
Moreover, a certain amount of internal interference (introduced by the computing platform) was also seen, which impacted on latency.
Since links in the VMLD were not totally independent, only partial benefits were obtained from diversity.
However, experiments highlighted a limited correlation between links, which means that in the real world MLO has a high capability to offer tangible improvements.

\bibliographystyle{IEEEtran}
\bibliography{bibliography}

\end{document}

%% file: tabs/stats_horizontal.tex
\small
\def\arraystretch{1.10}
\begin{tabular}{llrrrrrrrrrr}
\toprule
 & Set & Fraction \% & Average & Std. dev. & Min & $P_5$ & $P_{10}$ & $P_{95}$ & $P_{99}$ & $P_{99.9}$ & Max \\
\midrule
 & all & 100.0 & 322.1 & 835.8 & 97.8 & 107.4 & 108.0 & 1044.8 & 2781.8 & 13496.6 & 87490.4 \\
ch1 & acked & 92.7 & 315.2 & 795.7 & 97.8 & 107.4 & 108.0 & 1011.8 & 2625.4 & 12812.3 & 87490.4 \\
 & not acked & 7.3 & 409.3 & 1232.4 & 103.0 & 108.2 & 109.3 & 1613.1 & 4970.7 & 15526.8 & 54314.3 \\
\cline{1-12}
 & all & 100.0 & 390.1 & 780.5 & 99.8 & 107.9 & 108.5 & 1826.1 & 3645.4 & 6715.2 & 91672.4 \\
ch5 & acked & 94.0 & 397.2 & 778.0 & 99.8 & 107.8 & 108.4 & 1863.9 & 3665.5 & 6680.8 & 91672.4 \\
 & not acked & 6.0 & 279.3 & 810.7 & 99.8 & 109.2 & 115.6 & 754.5 & 3357.5 & 7205.1 & 75541.9 \\
\bottomrule
\end{tabular}